\documentclass[12pt,twocolumn,romanappendices]{IEEEtran}
\usepackage{cite}
\usepackage{graphicx, amsfonts,amsmath, enumerate, amsthm, color,amssymb,etaremune,array,url}
\usepackage{setspace,multirow,psfrag}
\usepackage{graphicx}
\usepackage{subcaption}
\usepackage{siunitx}
\graphicspath{{../figures/}}

\newcommand{\subparagraph}{}
 \usepackage[compact]{titlesec}
    \titlespacing{\section}{0pt}{2ex}{1ex}
    \titlespacing{\subsection}{0pt}{1ex}{0ex}
    \titlespacing{\subsubsection}{0pt}{0.5ex}{0ex}

\begin{document}
\title{The Road to 6G: Ten Physical Layer Challenges for Communications Engineers}
\author{Michail Matthaiou, Okan Yurduseven, Hien Quoc Ngo, David Morales-Jimenez, \\Simon L. Cotton, and Vincent F. Fusco

\thanks{
The authors are with the Institute of Electronics, Communications and Information Technology (ECIT), Queen's University, Belfast, Belfast BT3 9DT, U.K. (e-mail: m.matthaiou@qub.ac.uk).
}}
\maketitle

\begin{abstract}
While the deployment of 5G cellular systems will continue well into the next decade, much interest is already being generated towards technologies that will underlie its successor, 6G. Undeniably, 5G will have transformative impact on the way we live and communicate, yet, it is still far away from supporting the Internet-of-Everything (IoE), where upwards of a million devices per $\textrm{km}^3$ (both terrestrial and aerial) will require ubiquitous, reliable, low-latency connectivity. This article looks at some of the fundamental problems that pertain to key physical layer enablers for 6G. This includes highlighting challenges related to intelligent reflecting surfaces, cell-free massive MIMO and THz communications. Our analysis covers theoretical modeling challenges, hardware implementation issues and scalability, among others. The article concludes by delineating the critical role of signal processing in the new era for wireless communications.
\end{abstract}


\IEEEpeerreviewmaketitle
\vspace{-10pt}
\section{Introduction}
\color{black} After nearly eight years of intensive academic research and industrial testing on 5G, the lessons we have taken are the following: a) 5G can indeed support emerging data-hungry applications (e.g. ultra-fast broadband, high-definition video streaming), mainly through advances in the massive MIMO (mMIMO) space; b) 5G is still falling short of supporting the so-called Internet-of-Everything (IoE), where myriads of devices in a geographic cube require either low-latency, ultra-reliable connectivity or wireless Gpbs Internet access by availing of the mm-wave/THz spectrum \cite{Saad6G, Rappaport}. 
In the eve of a new decade, the idea of 6G has tentatively started to circulate within the wireless community, and the consensus is that
6G will try to address the shortcomings of 5G through \textbf{three scientific pillars} by boldly (1) pushing the communication to higher frequency bands (mm-wave and THz), (2) creating smart radio environments through reconfigurable surfaces and (3) by removing the conventional cell structures, aka cell-free massive MIMO. Yet, transforming these speculative academic concepts into commercially viable solutions is a very challenging exercise. Very recently, some related articles have elaborated on: 6G driving applications, metrics and new service classes \cite{Saad6G}, the importance and challenges of the THz-based communications \cite{Rappaport}, 
6G requirements and overview of supporting technologies \cite{Ming}, and, finally, potential use cases and 6G-enabling network architectures \cite{Zorzi}. 

On the other hand, our article moves away from the state-of-the-art and makes the following contributions: (i) we investigate the realizable potential of the three above mentioned scientific pillars; (ii) we identify {\textbf{ten immediate engineering challenges}} that need to be addressed \color{black} at the physical layer to boost follow-up research in the 6G ecosystem. We point out that the timescales of these different challenges vary depending on advances in other fields (e.g. electronics, CMOS design); (c) we finally articulate the critical role of fundamental signal processing (SP) in the 6G era along with the associated challenges. The vision of this article is to open up opportunities in electromagnetic theory, communication theory and transceiver design.

\color{black}

\vspace{-10pt}
\section{Intelligent Reflecting Surfaces (IRS)} \label{SectionII}
\textcolor{black}{The antenna architecture in a wireless communication system can be considered as the key front-end component, and will be the enabling physical technology for 6G networks. In the realization of the physical layer to facilitate the 6G technology, understanding the design challenges of the antenna structure is crucial to truly appreciate the potential of the next generation communication networks. In view of this, in this section, we aim to explain some of the key design challenges of IRS and show that the radiation characteristics of these apertures vary substantially as a result of these challenges.} 
A reflective surface is a planar aperture synthesized using an array of sub-wavelength elements (or unit cells). Due to their sub-wavelength unit cell sampling, reflective surfaces can be considered a distinct form of metasurfaces. There has been a substantial amount of research conducted in EM wave control using metasurface apertures with applications ranging from imaging to EM invisibility \cite{chen2016review}. Despite the fact that these structures are well understood within the applied EM community, their adoption in wireless communication networks has been \color{black} under-investigated \color{black} to date.  

Wireless systems conventionally rely on a mature antenna technology to establish a communication link, which is known as \textit{phased arrays}. A phased array consists of individual antennas with dedicated phased shifting circuits and power amplifiers to synthesize the desired aperture wavefront. Phased arrays can be power hungry and exhibit a rather complex hardware architecture. Different from the phased array technology, metasurfaces rely on a holographic principle to achieve the desired phase modulation. The incoming wave illuminating the aperture surface acts as a reference-wave, which is converted to a desired wavefront upon reflection from the reflective metasurface aperture. Their major advantage is that they can synthesize any arbitrary waveform using this simple, yet strong, holographic principle without the need for expensive and power hungry phase shifters. \textcolor{black}{Designing an IRS for wireless systems exhibits two main challenges: In Challenge 1, going beyond the conventional aperture level discussion, we consider the design challenges of the IRS considering their building blocks; i.e. the unit cells, and the effect of aberration on a unit cell level on the overall performance of the IRS. Satisfying the conditions of challenge 1, in challenge 2, we cover the dynamic reconfiguration aspect - a necessity for the "intelligent" operation, and review potential techniques to dynamically reconfigure the phase response of the unit cells.} 

\textit{\textbf{Challenge 1:} Unit cell phase range and phase quantization levels}

\begin{figure}[h!]
\centering
\includegraphics[width=7.5 cm,trim=0cm 0cm 0cm 0cm]{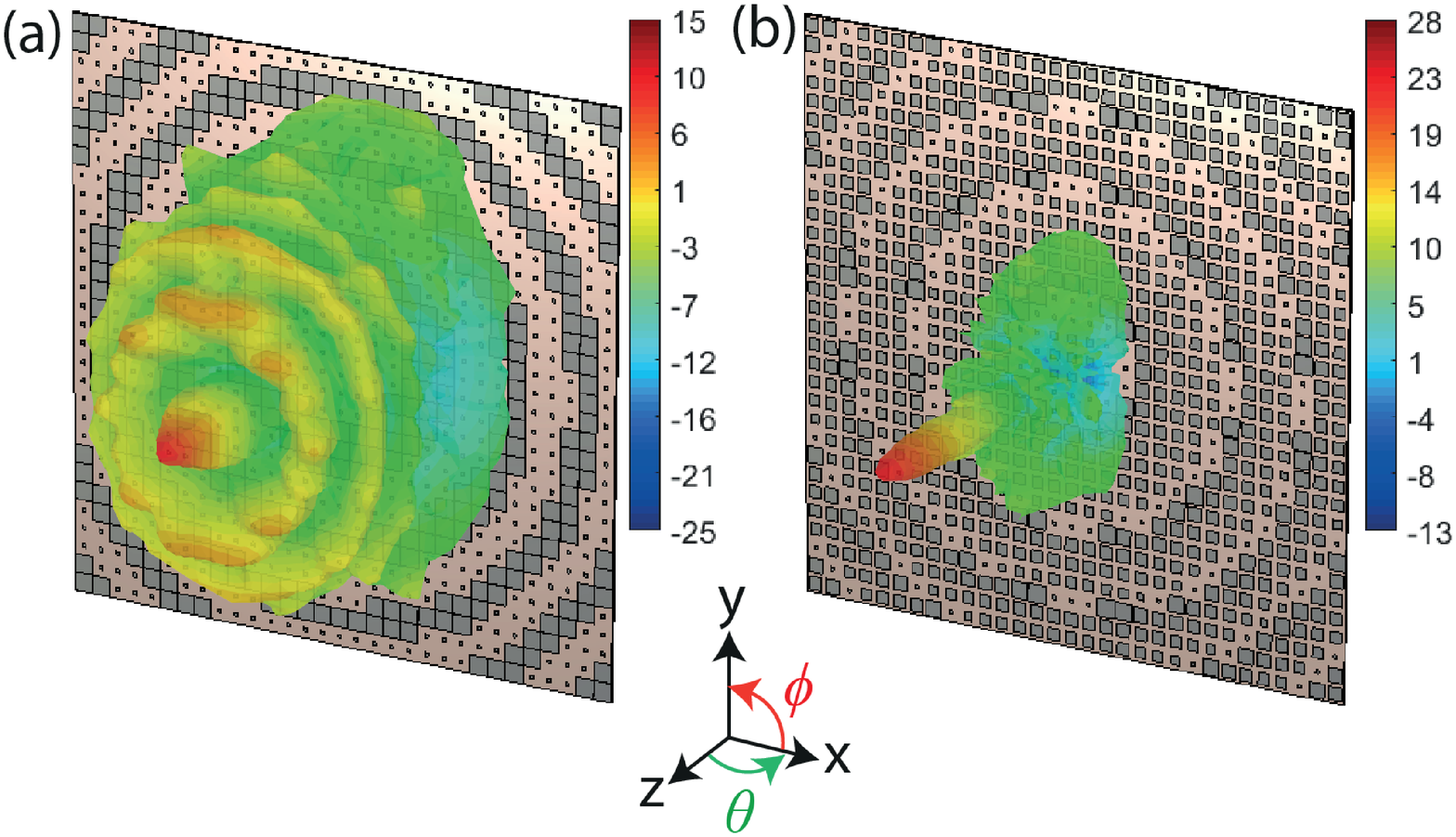}
\includegraphics[width=7 cm,trim=0cm 0cm 0cm 0cm]{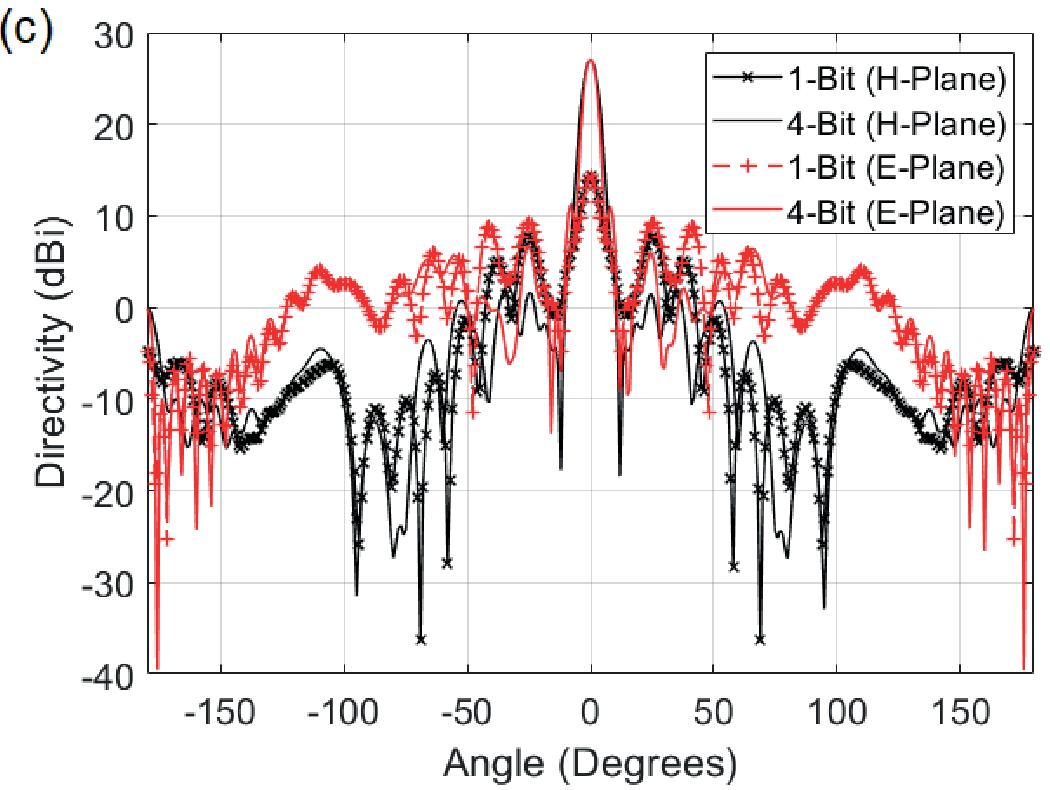}
\caption{\textcolor{black}{3D Radiation pattern of the reflective surface as a function of phase quantization level (a) 1-bit (b) 4-bit (c) comparison of the radiation patterns along the azimuth plane (H-plane) and elevation plane (E-plane). Colorbar in dB scale.}}
\label{fig:Figure_Sec_2_1}
\end{figure}

An important limitation in the design process of a reflective surface is the achievable phase range of the unit cells synthesizing the aperture. Ideally, each unit cell across the reflective surface should provide a full phase control across a phase range of 0-2$\pi$ radians. Another important constraint in the design process of a reflective surface is the quantization of the phase range of the unit cells. Even if a full phase range of 0-2$\pi$ radians is achieved, the number of quantization levels used to discretize this phase range has a direct impact on the fidelity of the synthesized wavefront upon reflection from the reflective surface. To investigate the effect of these design constraints, we consider a reflective metasurface and study these cases individually.
First, we assume that the unit cells can alter the phase response of the incoming reference-wave across the full phase range, 0-2$\pi$ radians. 
\color{black}Note that many examples appear in the literature for unit cell topologies that can achieve phase shifts of up to 2$\pi$, and the approaches used are now relatively mature, e.g. \cite{Venneri}. \color{black}
For this scenario, in Fig. \ref{fig:Figure_Sec_2_1}, we investigate two different quantization levels, 1-bit and 4-bit. We consider that the reflective metasurface is illuminated by an arbitrarily selected plane-wave incident along optical axis (z-axis) of the surface. The objective function is defined to be a wavefront radiated in the broadside direction ($\theta$=0$^{\circ}$, $\phi$=0$^{\circ}$) upon reflection from the surface. 

\begin{figure}[h!]
\centering
\includegraphics[width=7 cm,trim=0cm 0cm 0cm 0cm]{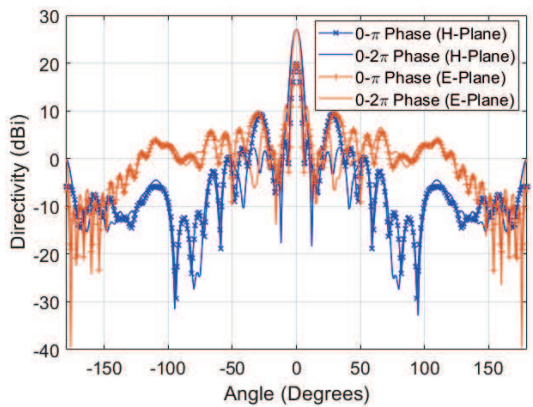}
\caption{\textcolor{black}{Radiation pattern of the reflective surface as a function of unit cell phase range.}}
\label{fig:Figure_Sec_2_2}
\end{figure}

\vspace{-2pt}
In Fig. \ref{fig:Figure_Sec_2_1}, the H-plane is the azimuth plane (or xz-plane) with $\phi$=0$^{\circ}$. Similarly, the E-plane is the elevation plane (or yz-plane) with $\phi$=90$^{\circ}$. Although, from a design perspective, using a 1-bit quantization level would be rather simple, in comparison to the 4-bit quantization case, the radiation pattern exhibits substantially higher sidelobes (8 dB higher) and less directivity (21 dB lower in the broadside direction). Increasing sidelobes is a significant concern for wireless communication networks due to the possibility to introduce strong interference across other channels. Similarly, reduced directivity adversely affects the link budget. As a result, from Fig. \ref{fig:Figure_Sec_2_1}, it is evident that there is a direct relationship between the phase quantization level and the fidelity of the radiation pattern of the reflective surface. \textcolor{black}{This is an important outcome because whereas the 1-bit quantization level can be achieved using a simple ON/OFF type binary modulation mechanism, phase quantization levels greater than that, such as the 4-bit scenario, require more sophisticated techniques to modulate the phase response of the unit cells, such as gray scale phase modulation \cite{chen2016review}. This can substantially increase the complexity of the unit cell design, and hence, a careful trade-off study is needed to achieve the desired hardware complexity and radiation performance from such apertures for 6G networks.} 
We now study the effect of the achievable unit cell phase range on the radiation pattern of the reflective surface. We select the quantization level to be maximum (4-bit) and investigate two unit cell phase ranges, 0-$\pi$ and 0-2$\pi$, respectively. The radiation pattern of the reflectarray surface vs the achievable unit cell phase range is shown in Fig. \ref{fig:Figure_Sec_2_2}.

From Fig. \ref{fig:Figure_Sec_2_2}, limiting the unit cell phase range to 0-$\pi$ reduces the directivity of the surface by 7.3 dB while increasing the sidelobe levels by as much as 7 dB. \textcolor{black}{This is an important outcome because whereas the 0-$\pi$ radians phase range can be achieved using a single resonant element, going beyond this phase range, such as the studied 0-2$\pi$ radians scenario, requires the combination of multiple independent resonances. As a result, careful consideration should be given to the achievable phase range from the unit cells to find the optimum trade-off between the complexity of the unit cell structure and the desired radiation characteristics of IRS apertures for 6G networks}. From Figs. \ref{fig:Figure_Sec_2_1} and \ref{fig:Figure_Sec_2_2}, it is evident that to have a realistic estimate of the link budget in a communication channel and a better understanding of interference characteristics, it is important that the design limitations of reflective surfaces are taken into account. Yet, it is often assumed that the reflective surface is ideal, suggesting that the link budget calculations do not consider these unit cell based aberrations. 

\textit{\textbf{Challenge 2:} Dynamic reconfigurability and IRS}

Albeit producing highly desirable radiation characteristics, conventional reflective surfaces are static, hence, beam characteristics are hard-coded into the surface during the design process. Different from static metasurfaces, an IRS has the capability to dynamically tune the reflection response of the aperture in an all-electronic manner. This is particularly important as communication environments have dynamic characteristics, namely variations in the number of connected users and non-static location distribution over time. Thus, the capability to intelligently change the characteristics of the reflected wavefront to meet their dynamic metrics plays a crucial role in future wireless communication systems. \textcolor{black}{The dynamic modulation of the IRS can be achieved using several techniques, such as leveraging materials with variable electrical properties, i.e. liquid crystals, or loading the unit cells with low-power semiconductor elements, e.g., PIN diodes and varactors \cite{yurduseven2018dynamically,chen2016review}}. \textcolor{black}{Challenges 1 and 2 are a direct consequence of an electromagnetic engineering problem. While both challenges can be addressed using rather complex, multi-layer, reconfigurable unit cell architectures \cite{chen2016review}, engineering such unit cells to exhibit low form-factor, loss, cost, system complexity and power consumption is a prerequisite in order for this technology to be a feasible option for future 6G networks.}  

\vspace{-3pt}
\section{Cell-free massive MIMO}
\begin{figure} \centering
    \begin{subfigure}[b]{\linewidth}    
        \includegraphics[width=0.9\textwidth]{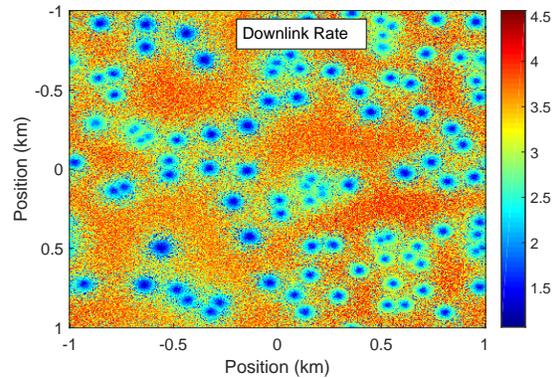} 
        \caption{Cell-free mMIMO} 
        \label{fig:cellfree}  
    \end{subfigure} 
    
    \begin{subfigure}[b]{\linewidth}
        \includegraphics[width=0.9\textwidth]{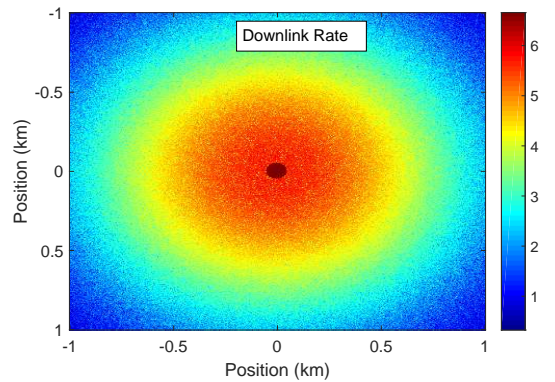}
        \caption{Colocated mMIMO}
                \label{fig:colocated}
    \end{subfigure} %
    \caption{\color{black} The downlink achievable rates for different user locations displayed with scaled colors (obtained using simulation approach described in  \cite{NAYLM:16:WCOM}); (a): $100$ APs are uniformly located at random in a $2\text{km}\times 2\text{km}$ area; (b): all 100 service antennas are located at the original point (i.e. the center of the square area).}
    \label{cell-freevscolo}
\end{figure}

\vspace{-3pt}
Cell-free mMIMO has been proposed in \cite{NAYLM:16:WCOM} to overcome the boundary effect of cellular networks. In cell-free mMIMO, many access points (APs) distributed in a geographic coverage area coherently serve many users in the same time-frequency resources. There are no cells, and hence, no boundary effects. Key points of cell-free mMIMO:
\begin{itemize}
\item \color{black} Cell-free mMIMO relies on mMIMO technology. More precisely, using many APs, cell-free mMIMO offers many degrees of freedom, high multiplexing gain, and high array gain. As a result, it can provide huge energy efficiency and spectral efficiency with simple SP.

\item In cell-free mMIMO, the service APs are distributed over the whole network, and hence, we can obtain macro-diversity gains. Thus, cell-free mMIMO can offer a very good network connectivity. There is no dead zone. Figures~\ref{fig:cellfree}--\ref{fig:colocated} show the downlink achievable rates displayed with scaled colors for cell-free mMIMO and colocated mMIMO, respectively. Clearly, cell-free mMIMO can offer much more uniform connectivity for all users.

\item Different \color{black} from \color{black} colocated mMIMO, where the base station is equipped with very large antennas, in cell-free mMIMO each AP has a few antennas. Thus, cell-free mMIMO is expected to be built by low-cost, low-power components and simple SP APs. 
\end{itemize}

The above benefits (in particular the high network connectivity) fulfil the main requirements of future wireless networks. Therefore, cell-free mMIMO has become one of the promising technologies of beyond 5G and towards 6G wireless networks, and attracted a lot of research interest \cite{interdonato2019ubiquitous}. {\color{black}Designing a low cost and scalable system is the ultimate objective of cell-free mMIMO research. To do this, we need scalable transmission protocols and power control techniques, which are discussed in Challenges 3 and 4. In addition, it is important to have new SP designs which can be implemented in a distribute manner to improve the system performance, scalability and robustness. This is discussed in Challenge 5. }


\color{black}
\textit{\textbf{Challenge 3:} Practical user-centric approaches}
\vspace{1pt}

In canonical cell-free mMIMO \cite{NAYLM:16:WCOM}, all APs participate in serving all users through the backhaul connections with one or several central processing units (CPUs). This is not scalable in the sense that it is not implementable when the network size (number of APs and/or number of users) grows large. Designing a scalable structure is one of the main challenges of cell-free mMIMO. It is shown in  \cite{NAYLM:16:WCOM} that, owing to the path loss, only 10-20\% of the total number of APs really participate in serving a given user.  Thus, each user should be served by a subset (not all) of APs. There are two ways to implement this: network-centric and user-centric approaches. In the network-centric approach, the APs are divided into disjoint clusters. The APs in a cluster coherently serve the users in their joint coverage area. Network-centric-based systems still have boundaries, and hence, are not suitable for cell-free mMIMO. By contrast, in user-centric approach, each user is served by its selected subset of APs. There are no boundaries, and hence, user-centric approach is a suitable way to implement cell-free mMIMO. There are several simple methods to implement user-centric approach, such as each user chooses some of its closest APs or chooses a subset of APs which contribute most of the total received power of the desired signal \cite{ngo2017total}.
 Yet, existing methods are not optimal, still require huge connections from all APs to the CPUs, {\color{black}and  are still fully network-controlled}. In addition, the cluster formed by each user changes quickly depending on the user locations. This requires more control signaling. Hence, designing a practical user-centric approach is a  challenging research exercise.
 
\textit{\textbf{Challenge 4:} Scalable power control}

Power control is central in cell-free mMIMO since it controls the near-far effects and the interuser interference to optimize the objectives (e.g. the max-min fairness or the total energy efficiency) we want to achieve. Ideally, power control is done at the CPU under the assumption that the CPU  perfectly knows all large-scale fading coefficients. Then, the optimal power control coefficients will be sent to the APs (for the downlink transmission) and to the users (for the uplink transmission). This requires huge front/back-hauling overhead. Yet, it is very difficult for the CPU to have perfect knowledge of large-scale fading coefficients associated with a potentially unprecedented number of APs and users. {\color{black}Thus, besides the unscalability of the current canonical transmission protocol (discussed in Challenge 3), the above power control method also creates another issue which makes the system unscalable.} Thus, power control should be done distributed at the APs with local knowledge of the channel conditions. This is again problematic because it is hard to control the near-far effects and interuser interference without full channel knowledge of all links from all APs and users. Some heuristic power control schemes have been proposed, which, however, are developed based on a specific assumption of the propagation environment, and hence, it is hard to evaluate how well these schemes work in practice \cite{interdonato2019ubiquitous}.
Promising approaches based on machine learning (ML) and deep learning (DL) have recently been proposed \cite{Chen2019}. A key question is whether these approaches are also scalable, in order to meet the foreseeable decentralization of  cell-free mMIMO.

\vspace{1pt}
\textit{\textbf{Challenge 5:} Advanced distributed SP}
\vspace{1pt}

One of the ultimate aims of cell-free mMIMO research is designing a SP scheme which offers good performance and can be implemented in a distributed manner. Otherwise, the system will not be scalable. In canonical cell-free mMIMO \cite{NAYLM:16:WCOM}, \color{black} conjugate beamforming is normally considered since it can be implemented in distributed manner and performs well\color{black}. Yet, compared to other linear processing schemes, such as zero-forcing and minimum mean-square error, the performance of conjugate beamforming is far below. To cover the gap between conjugate beamforming and ZF/MMSE, we need to have additionally very large number of service antennas. Cell-free mMIMO with local ZF was proposed in \cite{interdonato2019local}. However, this scheme requires that each AP has a large number of antennas. This is more challenging for the uplink design. Currently, there are no distributed SP schemes available for the uplink. Even with the simple matched filtering, we need to send the (processed) signals from each AP to the CPUs for signal detection.

\vspace{-5pt}
\section{Moving to higher frequency bands}
6G wireless systems  will rely on:
i)	Millimeter-wave technologies (30 to 300 GHz); ii) THz technologies (300 GHz to 3 THz); and
iii) Free space optics (FSO) \cite{Rappaport}. 
\color{black} It is known that the exacerbated atmospheric attenuation and path loss at higher frequencies, can be compensated by miniaturized massive antenna arrays that can support super sharp beamforming. Also, there are natural synergies between IRS/cell-free mMIMO and higher frequency systems to extend the operating range and combat the fundamental distance problem. We now overview three  challenges towards this objective.

\color{black}

\vspace{1pt}
\textit{\textbf{Challenge 6:} Packaging/interconnect techniques }
\vspace{1pt}
In addition to providing for a physical enclosure, packaging must provide a reliable interconnection between interior and exterior operating environments. Some of the key driving factors include integration of high-speed semiconductor integrated circuits with advanced antenna systems and integration with optoelectronics. 
At higher frequencies, bond wires cause considerable signal degradation. The effects of bond wires are difficult to characterize for large signal applications, such as power amplifiers and also for phase critical applications, such as beamformers for phased arrays, where these can introduce side lobe levels. Traditional metallic split-block packages provide excellent performance but are bulky and heavy. Noticeable progress has been made in interconnect and packaging technologies for THz applications. Cutting-edge techniques in micro-machining and LTCC technology yield compact and low-cost solutions. Additive manufacturing techniques, such as metal coated 3D printing of plastic devices, can realize low-cost, light weight and compact devices. \color{black}Ceramic packages for both low- and high-power devices are commercially available for applications up to, 50 GHz. \color{black}

\vspace{1pt}
\textit{\textbf{Challenge 7:} Transceiver design}
\vspace{1pt}

The compact physical size and power efficiency requirements become more challenging at higher frequencies. Hybrid beamforming will be best suited to implement large number of antenna elements along with high efficiency amplifiers. Performance parameters, such as the noise figure, output power and power efficiency degrade significantly at high frequencies. Demodulation of higher order modulated signals also becomes more challenging as phase noise increases at higher frequencies. Advanced array SP techniques must complement the transceiver design to address these challenges. Novel techniques, such as spatially oversampled antennas and new phased array architectures can be leveraged to provide a solution to size, weight and power consumption of large mm-wave/THz antenna arrays. \color{black} MMIC beamformers and compact millimeter wave wireless communications modules incorporating high gain phased antenna arrays that are capable of achieving high-speed data transfer at frequencies up to 60 GHz are presently commercially available. \color{black} 

\vspace{-13pt}
\textit{\textbf{Challenge 8:} Measurements \& standardization}
Two popular approaches for phase-sensitive measurements at THz use either vector network analyzers (VNAs) or time domain spectrometers \cite{THz1}. In both cases, calibration, verification, and measurement traceability at THz frequency bands remains a major challenge. For time domain systems, a major challenge is the establishment of standardized measurement and calibration, whereas for VNA systems, solutions are being sought for high precision waveguides and interconnects. Electro-optic sampling is promising as a complementary approach to THz measurements, though it is yet to extend the bandwidth to 1.5 THz and to improve resolution. \color{black} 

\vspace{-3pt}
\section{The Role of SP in the 6G Era}
\vspace{-3pt}
The journey towards 6G will inevitably be hurdled by significant challenges in \color{black} the SP arena. \color{black} Massively populated and decentralized (cell-free) networks, supporting unprecedented Internet of Everything connectivity, will produce high-dimensional and increasingly complex signals, which are subject to increased interference and other impairments (e.g., related to synchronization, temporal correlation etc.) that have so far been largely overlooked. Current SP methods, generally based on low-dimensional signals and classic stationarity assumptions, will need to be rethought. We now discuss two of these methods: channel estimation and adaptive filtering, and their associated challenges.

\textit{\textbf{Challenge 9:} Channel estimation}

Extensive research in the context of 5G has been on how to reduce training overheads in pilot-based channel estimation. While such need will be most critical in 6G, viable solutions become extremely challenging due to the massive scale-up and connectivity demands, in particular: (i) supporting high data rates (Gbps) in high-mobility scenarios---a prime concern of operators---will require dealing with much shorter channel coherence times; (ii) ultra-low latency requirements will see transmission intervals substantially shortened, and (iii) the number of parameters to estimate will be massively large as a consequence of the scaling (not only of antennas/APs, but also of users/devices). The high-dimensional channels will need to be estimated in severe under-sampling constraints, which might render pilot-based (coherent) estimation unfeasible, particularly under high-mobility or low-latency requirements. Blind (non-coherent) estimation approaches---which do not require dedicated pilot signals---stand as promising alternatives. {\color{black} Though significant efforts have been made in this direction, existing approaches require knowledge of the (high-dimensional) signal covariance matrix \cite{Ngo2012}}, which will again need to be acquired from a limited number of samples. To that end, {\color{black} fundamental research in the fields of random matrix theory (RMT) and high-dimensional statistics will be crucial}; in particular, to develop accurate estimators of large covariance matrices (and their eigen-spectrum) under limited sampling. More recent approaches based on ML might also play a relevant role \cite{Chen2019}.


\vspace{1pt}
\textit{\textbf{Challenge 10:} Adaptive filtering}
\vspace{1pt}

In beamforming, transmitted signals are dynamically adapted (via digital precoding) to the propagation conditions, effectively mitigating interference and noise. Adaptive beamforming can be seen as a linear filter with a particular design objective, e.g., to maximize the signal-to-noise ratio (SNR). Optimal solutions for the beamformer (and associated receiver filters) require the covariance matrix of the aggregated interference and noise; unknown in practice, this needs to be estimated from observed samples. Current solutions rely on classical estimators, such as the sample covariance matrix (SCM), which will return a poor estimate in high-dimensional 6G scenarios, due to:
\vspace{-3pt}
\begin{itemize}
\item \textit{Scarcity of samples}: While the numbers of antennas and user devices will scale up massively, strict low-latency and high-mobility requirements will impose a rather limited number of training (observed) samples.

\item \textit{Temporal correlation}: The ultra-dense and highly decentralized deployments, e.g., with thousands of distributed APs in cell-free networks, will be subject to non-perfect synchronization (e.g., between interference and desired signals) and non-stationarity effects.

\item \textit{Outlying samples}: With millions of interconnected devices (from electrical/smart appliances to connected vehicles), we expect multiple sources of impulsive noise and, in security-sensitive applications, eventual sources of intentional interference (jamming).
\end{itemize}
\vspace{-3pt}
Traditional estimators (e.g., SCM) rely on the sufficient availability of samples (the number of samples should be far greater than the number of signals) and on the assumption that these samples are i.i.d. Under the conditions above, however, the mismatch between true and estimated covariance leads to highly inaccurate filters with severe performance losses, in terms of connectivity, reliability, and data rates. A {\color{black} fundamental} challenge is then to develop filtering solutions which are robust to the effects mentioned above. {\color{black} While research to that end has so far been very scarce, promising tools and directions can be leveraged from the fields of robust statistics, RMT, and high-dimensional covariance estimation (see, e.g. \cite{Auguin2018} and references therein)}.




\vspace{-8pt}
\section{Conclusion}
\vspace{-3pt}
Although the 6G era is a decade away, it is extremely timely to understand what are the main challenges for communications engineers. We identified \color{black} ten immediate  challenges \color{black}whose investigation will cross-leverage expertise in SP, information theory, electromagnetics and physical implementation. 

\bibliographystyle{IEEEtran}


\vspace{-3pt}

\section*{Authors}
\vspace{-3pt}

\begin{small}

{\bf\em {Michail Matthaiou}} is a Professor at Queen's University Belfast (QUB), UK. His research spans signal processing, massive MIMO, and mm-wave systems.

{\bf\em {{Okan Yurduseven}}} is currently a Senior Lecturer at QUB. His research interests include microwave and mm-wave imaging, MIMO radar, antennas and metamaterials. 

{\bf\em {{Hien Quoc Ngo}}} is a Lecturer at QUB. His main research interests include massive MIMO, cell-free massive MIMO, physical layer security, and cooperative communications. 

{\bf\em {{David Morales-Jimenez}}} is a Lecturer at QUB. 
His research interests span statistical signal processing, random matrix theory, and high-dimensional statistics.

{\bf\em {{Simon L. Cotton}}} is a Professor of Wireless Communications at QUB. Among his research interests are propagation measurements and statistical channel modeling.

{\bf\em {{Vincent F. Fusco}}} is the Chair of High Frequency Electronic Engineering at QUB. His research interests include active antenna front-end techniques and self-tracking antenna arrays.

\end{small}
\end{document}